\documentclass[aps,prb,twocolumn,showpacs,showkeys,amsmath,amssymb]{revtex4-2}
\usepackage[colorlinks=true,citecolor=blue]{hyperref}

\def\fileversion{\it version~1.0.0%
(\if\pdfstrcmp{\pdffilesize{\jobname.tex}}{32592}0{original}\else{the file was changed!}\fi)}

\DeclareMathOperator{\ccirc}{\mathrel{\circ\kern-.91ex\raise-0.3ex\hbox{,}}}

\begin{document}

\title{Dirac quantum kinetic equation: minimal conductivity revisited}
\author{Oleksiy Kashuba}
\email[Email: ]{o.kashuba@gmail.com}
\author{Bj\"orn Trauzettel}
\affiliation{Theoretische Physik IV, Institut f\"ur Theoretische Physik und Astrophysik, Universit\"at W\"urzburg, 97074 W\"urzburg, Germany}
\author{Laurens W.\ Molenkamp}
\affiliation{Experimentelle Physik III, Physikalisches Institut, Universit\"at W\"urzburg, 97074 W\"urzburg, Germany}
\date{\today }

\begin{abstract}
The kinetic equation used for the description of Dirac systems does not fully take into account two features that play an important role in the vicinity of the Dirac point: (i) the spin degree of freedom, in particular if the spin-flip energy $2 vp$ is not large anymore; and (ii) the failure of the semiclassical approximation due to the large Fermi wavelength.
In our work, we propose a novel quantum kinetic equation, which does not have these two drawbacks.
Exploiting it in the presence of short range disorder, we demonstrate how it predicts the correct minimal conductivity in 2D Dirac system, a result that has so far been obtained only by other methods like the Kubo formula.
The nature of the presented kinetic equation opens up the possibility for the kinetic description of deeply quantum and even strongly correlated Dirac systems.
\end{abstract}

\maketitle

\section{Introduction}

In last years, the scientific community constantly discovered new materials, which posses distinct features such as linear dispersion and strong spin-orbit coupling described by effective Dirac-like Hamiltonians~\cite{Balatsky2014}.
Graphene was the first 2D material that revealed the linear dispersion in the Hamiltonian operating in isospin (lattice site) space~\cite{CastroNeto2009}.
Similar spectra were found in 2D surface states of 3D topological insulators~\cite{Hasan2010,Qi2011}, Weyl or Dirac semimetals~\cite{Armitage2018}, superconductors with $d$-wave pairing~\cite{Sun2015}, and liquid $^{3}$He~\cite{Volovik1992}.
These materials share a common theoretical basis.
The kinetic equation, despite its simplicity, is a powerful theoretical instrument for the description of the behavior of a given system on scales much larger then the internal scales.
To describe the non-equilibrium dynamics in Dirac systems, the Boltzmann equation has been adapted by introducing two distribution functions corresponding to the positive and negative energies of the Dirac cones.
This description neglects the spin degree of freedom but takes helicity into account~\cite{Shon1998,Auslender2007}.
Nevertheless, it has been successfully applied for the calculation of the transport properties as for dominating disorder~\cite{Ando2006,Adam2007,Hwang2007,Kechedzhi2008,Katsnelson2008}, so for strong electron-electron scattering, leading to hydrodynamic transport~\cite{Fritz2008,Muller2008b,Kashuba2018}.
This approach is generally validated by the magnitude of the momentum accompanied by strong spin-momentum locking.
The Berry phase contributions have been ignored until the wave package approach was developed~\cite{Niu1999,Niu2004,Niu2006}, which allowed to take the torque dipole contribution into account.
An alternative formalism based on preserving the original spin matrix structure in the kinetic equation has recently been proposed~\cite{Kashuba2019}.
It demonstrates the out-of-plane deflection of the spin in 2D and predicts a transverse spin current.


All methods mentioned above have a substantial drawback---they operate with the distribution function, or more generally, the semiclassical Green function~\cite{RammerSmith1986,Rammer2007}, used in all parts of the kinetic equation including the collision integral.
This self-consistency breaks down in the vicinity of the Dirac point, where the Fermi wavelength is not the shortest scale, and the mass surface is not well-defined anymore.
The kinetic equations written so far were not able to reproduce the well-known expression for the minimal conductivity in graphene ($e^{2}/\pi h$ per channel).
We significantly extend the ideas expounded in Ref.~\onlinecite{Kashuba2019}, abandoning the semiclassical approximation.
The presented formalism does not require large Fermi momentum, or sharp resonance in the frequency dependence of the spectral function.

We start from the Dyson equations containing all quantum information of the system.
Using a general ansatz~\cite{RammerSmith1986}, we separate the equations describing the spectral properties, and arrive at a single equation for the generalized distribution function.
We refer to this equation as the generalized quantum kinetic equation.
We simplify it for the case of short range disorder, and solve it calculating the linear response to an applied voltage.
This allows us to obtain the correct result for the conductivity in the charge neutral case~\cite{Ludwig1994,Ziegler2007}.

The article is organised as follows.
In Section~\ref{sec:qke} we introduce generalized distribution function and present the quantum Dirac kinetic equation.
We derive the collision integral for short-range disorder and an expression for the current.
Next Section~\ref{sec:cond} presents the details of the kinetic equation solution in the presence of the small voltage bias (i.e.\ chemical potential gradient) and gives an expression for the conductivity of the investigated Dirac system.

\section{Quantum kinetic equation}
\label{sec:qke}

In this section, we derive the general quantum kinetic equation for spin-orbit coupled systems and simplify it for the case of short range disorder.
The Keldysh formalism operates with three Green functions: retarded/advanced components, which are complex conjugates and can be obtained solely by performing the analytic continuation of the spectral function $A=i(G^{R}-G^{A})$ to the upper/lower half-plane, and the Keldysh component, which contains the information about the occupation of the quantum states~\cite{RammerSmith1986}.
The spectrum and the occupation parts can be separated using an ansatz based on the generalized distribution function $f$ (see Ref.~\onlinecite{RammerSmith1986} and Appendix~\ref{apx:qkederiv}).
This distribution function depends on two times and two coordinates, but unlike the canonical distribution function, one needs to combine it with the spectral function in order to obtain the physical correlator:
\begin{equation}
\langle \psi_{1'}^{+}\psi_{1}\rangle =\{A \ccirc f\}_{11'}
\equiv
\int_{2\leq1}A_{12}f_{21'}+\int_{2\leq1'}f_{12}A_{21'}.
\end{equation}
The subscripts denote time, coordinates and spin: $1\equiv (t,\mathbf{r},\alpha)$, $1'\equiv (t',\mathbf{r}',\beta)$, and $2\equiv (t_{2},\mathbf{r}_{2},\gamma)$.
The spectral function and integrals are defined as
\begin{equation}
A_{11'}= \langle  \psi_{1}\psi^{+}_{1'} + \psi^{+}_{1'}\psi_{1}\rangle,
\quad
\int_{2\leq1}\equiv \int_{t_{2}\leq t}d^{d}\mathbf{r}_{2}dt_{2},
\label{eq:Aintdef}
\end{equation}
where $d=2$ or $3$ is the dimensionality of the system.
Thus, the brackets are equivalent to $\{A \ccirc f\} \equiv i(G^{R}\circ f- f \circ G^{A}) = i G^{<}$, where the circle denotes the ordinary convolution, i.e.\ an integration like in Eq.~\eqref{eq:Aintdef}, but over all values of $t_{2}$.
All six Dyson equations (left and right for all three components) result in one single equation for the generalized distribution function $f$ (see Ref.~\onlinecite{RammerSmith1986} and Appendix~\ref{apx:qkederiv}):
\begin{equation}
\frac12\{\mathcal{D}\ccirc f\} = K -\frac12 \{\varGamma\ccirc f\}.
\label{eq:genkineq}
\end{equation}
This is a matrix equation in $2\times2$ spin space, with each component depending on two spin indices, two times, and two coordinates.
The right part of this equation, where $\mathcal{D}_{11'}=(\partial_{t_{1}}+iH_{1})\delta_{11'}$, and $H_{1}$ is the bare Hamiltonian, contains the kinetic terms. 
The left part contains the self-energy contributions: $K=i\varSigma^{<}$, $\varGamma=i(\varSigma^{R}-\varSigma^{A})$, and $\{\varGamma \ccirc f\} \equiv i(\varSigma^{R}\circ f- f \circ \varSigma^{A})$.
The net expression in the left hand side of our quantum kinetic equation plays the same role as the collision integral in the conventional Boltzmann equation.
Our equation, however, is exact and contains all quantum mechanical details.

It is better to perform further calculations using the mixed representation---when the Fourier transform is taken over the difference of the temporal/spatial variables only.
In this case, the generalized distribution function depends on time, coordinate, frequency, and momentum: $f=f_{\alpha\beta}(t,\mathbf{r},\omega,\mathbf{p})$.
The convolution in this case can be represented in the form of the derivatives over all four arguments (see Appendix~\ref{apx:qkederiv}).
In contrast to the conventional distribution function, the generalized one additionally depends on the frequency $\omega$, containing thus, information about real and virtual processes.

The use of this generalized distribution function has several advantages.
In Dirac systems with chemical potential close to the neutrality point the mass surface is not well-pronounced.
Therefore, we cannot write self-consistent equations containing only the semiclassical (integrated by frequency or energy) Green function.
Instead, we must use the original lesser Green function $G^{<}$.
However, it has a significant drawback: a strong dependence on frequency.
The generalized distribution function $f$, in turn, is a smooth function of both frequency and momentum.

In the absence of electromagnetic fields, the Dirac Hamiltonian is $H_{1}\equiv -i\boldsymbol{\sigma}\cdot\partial_{\mathbf{r}_{1}}$, so the gradient part of the kinetic equation~\eqref{eq:genkineq} becomes
\begin{equation}
\partial_{t} f + \frac12 \partial_{\mathbf{r}} \cdot [\boldsymbol{\sigma},f]_{+}
+ i \mathbf{p}[\boldsymbol{\sigma},f]_{-} = 
K -\frac12 \{\varGamma\ccirc f\}.
\label{eq:kegeneral}
\end{equation}
This is the central equation of our paper.
Despite its simplicity, it has a number of advantages with respect to other approaches:
(1) It is exact, not a gradient expansion, and fully reproduces results of the Dyson equation.
(2) Contrary to the Dyson equation, it is a single kinetic equation, the solution of which (a) complies to both left and right equations for the Keldysh component, (b) is a smooth function unlike the Green functions.
Such benefits are achieved by the price of adding one more argument (frequency $\omega$) to the distribution function, which is a $2\times2$ matrix in spin space.
As we show below, the solution of this equation is not a complicated task.

To obtain the right part of Eq.~\eqref{eq:kegeneral}, which represents the collision integral, we consider (as an example) the scattering off impurities with the Fourier-transformed potential $U_{\mathbf{p}}$. 
In Born approximation, we get
\begin{align}
\varGamma(t,\mathbf{r},\omega,\mathbf{p}) &= \int \frac{d^{2}\mathbf{p}'}{(2\pi\hbar)^{2}}  w_{\mathbf{p}-\mathbf{p}'} A(t,\mathbf{r},\omega,\mathbf{p}'),
\\
K(t,\mathbf{r},\omega,\mathbf{p}) &= \int \frac{d^{2}\mathbf{p}'}{(2\pi\hbar)^{2}}  w_{\mathbf{p}-\mathbf{p}'} \frac12\{A\ccirc f\}(t,\mathbf{r},\omega,\mathbf{p}'),
\end{align}
where $w_{\mathbf{p}} = n_{i}|U_{\mathbf{p}}|^{2}$ and $n_{i}$ is the density of the impurities.

In real physical (Dirac) systems, charged impurities are often the main source of disorder.
The screened Coulomb shape of the potential $U_{\mathbf{p}}$ created by separate impurities has been believed to be responsible for the finite value of the minimal conductivity in graphene~\cite{Nomura2007}, but later it was demonstrated by means of the Kubo formula, that short-range disorder can also lead to the finite value of the minimal conductivity~\cite{Ludwig1994,Ziegler2007}.
The scattering amplitude can be approximated as $w_{\mathbf{p}} = w$ for $vp<E$ and $w_{\mathbf{p}} = 0$ for $vp>E$, where the cutoff $E$ is large compared to all other energy scales.
In this case, the scattering rate depends on frequency only and looses its spin structure: $\varGamma_{\alpha\beta}=\delta_{\alpha\beta}\Gamma(\omega)$~\footnote{%
We use the italic font for the matrices in the spin space and roman font for the degenerate spin dependence (see Eq.~\eqref{eq:hmatrixparam} for example).
}.
The spectral function can be directly expressed through $\Gamma$ via:
\begin{equation}
A = \sum_{\lambda} \frac{1+\boldsymbol{\sigma}\cdot \mathbf{n}_{\lambda}}{2} \frac{\Gamma}{(\omega- \lambda v p )^{2} + \Gamma^{2}/4 },
\end{equation}
where $ \mathbf{n}_{\lambda}= \lambda\mathbf{p}/p$ and $\lambda = \pm$ designates helicity, i.e.\ the upper/lower Dirac cone.
Defining the angle brackets as
\begin{equation}
\langle \ldots \rangle = \frac{w}{2(2\pi\hbar)^{d}}\sum_{\lambda}\int_{vp<E}\ldots \frac{d^{d}\mathbf{p}}{(\omega-\lambda vp)^{2}+(\Gamma/2)^{2}},
\label{eq:bracketsdef}
\end{equation}
we can write the self-consistent equation for $\Gamma$: $\Gamma=\langle \Gamma \rangle$, which can be reduced to $1=\langle 1 \rangle$ while $\Gamma$ depends on frequency only.
In the 2D case, for example, the integration gives us a known result:
\begin{equation}
\Gamma(\omega) = 
\begin{cases}
\Gamma_{0} & \omega \ll \Gamma_{0}, \\
\pi\alpha\omega & \omega \gg \Gamma_{0},
\end{cases}
\end{equation}
where $\Gamma_{0}=2E e^{{-1/\alpha}}$ and the dimensionless parameter $\alpha = w/2\pi\hbar^{2}v^{2}$ is typically small~\cite{Mirlin2006}.
The details of this calculation can be found in Appendix~\ref{apx:sezcalc}.

In linear response, the temporal $t$ and spatial $\mathbf{r}$ scales are much larger than the scales involved in the Fourier transform, namely $1/\omega$ and $1/p$.
This justifies the convolution expansion over  frequency and momentum derivatives.
On the right hand side of Eq.~\eqref{eq:kegeneral}, we take the zero order only, $\varGamma \circ f \approx \varGamma f$, similar to the canonical kinetic equation.
For short-range disorder, the self-energy becomes spin degenerate, and the real part $\varSigma^{R}+\varSigma^{A}$ vanishes.
Note that for short-range disorder the self-energy does not depend on momentum, while, in the static case, $f$ does not depend on time.
If both assumptions apply, the convolution is exactly equal to the product.
After resolving the spin structure of all components
\begin{equation}
K= \mathrm{K} + \boldsymbol{\sigma}\cdot \mathbf{K},
\qquad
\varGamma=\Gamma,
\qquad
f= \mathrm{f} + \boldsymbol{\sigma}\cdot \mathbf{f},
\label{eq:hmatrixparam}
\end{equation}
we get a set of kinetic equations for scalar $\mathrm{f}$ and vector $\mathbf{f}$ components of the generalized distribution function:
\begin{equation}
\begin{split}
\partial_{t}\mathbf{f} + v \nabla \mathrm{f} -2v \mathbf{p} \times \mathbf{f} &= \mathbf{K} -\Gamma \mathbf{f},
\\
\partial_{t}\mathrm{f} +v \nabla \cdot \mathbf{f} &= \mathrm{K} -\Gamma \mathrm{f}
\end{split}
\label{eq:qkespinresolved}
\end{equation}
with the components of the right part of the Eq.~\eqref{eq:kegeneral} equal to
\begin{equation}
\begin{split}
\mathrm{K} &= \Gamma \langle \mathrm{f}+ \mathbf{n}_{\lambda} \cdot \mathbf{f}\rangle,
\\
\mathbf{K} &= \Gamma \langle  \mathbf{n}_{\lambda} \mathrm{f}+ \mathbf{f}\rangle - 2\langle (\omega-\xi)\mathbf{n}_{\lambda} \times \mathbf{f}\rangle.
\end{split}
\label{eq:Kdef}
\end{equation}
Here, $\xi = \lambda v p$, and both $\mathrm{K}$ and $\mathbf{K}$ do not depend on momentum.
These equations have two distinct features: 
(1) the vector part $\mathbf{f}$ of the distribution function, which captures the spin distribution; 
(2) the term $\mathbf{p} \times \mathbf{f}$, which for large momenta ($vp \gg \Gamma$) guarantees strong spin-momentum locking.
The vector product in this term also exists in the 2D systems.
It is responsible for the generation of the off-plane $z$-component of the spin polarization~\cite{Kashuba2019}.

The short-range approximation for disorder has an additional advantage: we do not have to calculate the current from the distribution function, but can do it on basis of the vector part $\mathbf{K}$ of the collision integral:
\begin{multline}
\mathbf{j} = ev \iint\frac{d\omega d^{2}\mathbf{p}}{(2\pi\hbar)^{3}}\mathrm{Tr}[\boldsymbol{\sigma}(-iG^{<})]
=\\= - \frac{ev}{w}\int\frac{d\omega }{2\pi\hbar} \mathbf{K} = - \frac{e^{2}/h}{2\pi\hbar^{2}v}\,\frac{1}{e\alpha}\int \mathbf{K} d\omega.
\label{eq:currentdef}
\end{multline}
where $h/e^{2}\approx26 \,\mathrm{k}\Omega$ is the von-Klitzing constant.


\section{Linear response to the gradient of chemical potential}
\label{sec:cond}

To demonstrate the superiority of the quantum kinetic equation~\eqref{eq:qkespinresolved} we solve it for the static case calculating the conductivity for the charge neutral case.
The charge transport is generally invoked by the weak gradient of the chemical potential $\nabla \mu$ creating a voltage drop $\Delta V = \frac1e \int \nabla \mu \cdot \mathbf{r}$.
In conventional metals the gradient of the chemical potential and electric field are often used as synonyms, but general this is not correct.
One of the manifestations of their difference is Klein tunnelling through electrostatic potentials, which does not occur for a non-uniform distribution of electrons in a Dirac material.

In the static case the Eq.~\eqref{eq:qkespinresolved} gives us an expression for the vector part of the distribution function
\begin{equation*}
\mathbf{f} = \Gamma^{-1}\hat{C}(\mathbf{K}-v\nabla \mathrm{f}),
\quad
\hat{C}\mathbf{X} = \frac{\mathbf{X}+\mathbf{q}\times\mathbf{X}+ \mathbf{q}(\mathbf{q}\cdot\mathbf{X})}{1+\mathbf{q}^{2}},
\end{equation*}
where $\mathbf{q}=2v\mathbf{p}/\Gamma$.
Substituting it into Eq.~\eqref{eq:Kdef}, we get
\begin{equation*}
\mathbf{K} = \Gamma \langle \mathbf{n}_{\lambda} \mathrm{f}\rangle + \bigl\langle \hat{Q}(\mathbf{K}-v\nabla \mathrm{f})\bigr\rangle,
\quad
\hat{Q} = \hat{C} + (1-\omega/\xi)(\hat{C}-1).
\end{equation*}
In case of weak nonequilibrium (i.e.\ linear response), we can perform the expansion in gradients, namely $\Gamma^{-1}v\nabla$.
In zeroth order expansion, $f$ is close to the equilibrium Fermi distribution, i.e.\ $f^{(0)}=\mathrm{f}^{(0)} = \mathrm{f}_{F}\bigl(\omega-\mu(\mathbf{r})\bigr)$.
In first order expansion, we obtain a value for $\mathbf{f}$, as stated above, but unlike in the conventional kinetic equation, the value $\langle \mathbf{n}_{\lambda} \mathrm{f}\rangle = \Gamma^{-1}v \langle\mathbf{n}_{\lambda} \nabla \mathbf{f}\rangle$ is of second order in the gradient expansion.
Therefore, the momentum dependence of the scalar part $\mathrm{f}$ can be neglected.
This is not surprising if we recall that the current in Dirac systems is not the concentration of the electrons with an average momentum polarization, but rather the aggregation of the electrons with a spin polarization~\footnote{%
In the systems with conventional $\mathbf{p}^{2}$ Hamiltonian the current is proportional to $\sum_{\alpha\mathbf{p}}\mathbf{p}c_{\alpha\mathbf{p}}^{+}c_{\alpha\mathbf{p}}$.
In Dirac systems the current is $\sum_{\alpha\beta\mathbf{p}}c_{\alpha\mathbf{p}}^{+}\boldsymbol{\sigma}_{\alpha\beta}c_{\beta\mathbf{p}}$, which coincides with the other definition only in the equilibrium case.}.
This leads us to the self-consistent equation $\mathbf{K} = \langle \hat{Q}\rangle(\mathbf{K}-v\nabla \mathrm{f})$, solved as
\begin{equation}
\!\!\!\!\mathbf{K} = \left(\!1\!+\!\frac{d/2}{d-1}\frac{1}{Z-1}\!\right)v\nabla \mathrm{f},
\quad
Z=\left< \!\frac{2\lambda\omega vp+\Gamma^{2}}{4v^{2}p^{2}+\Gamma^{2}}\!\right>.
\label{eq:KcalcZdef}
\end{equation}
For the 2D case, the expressions take the form (see Appendix~\ref{apx:sezcalc} for details)
\begin{equation*}
\mathbf{K} = \frac{Z}{Z-1}v\nabla \mathrm{f},
\quad
\mathbf{f} = \frac{\Gamma^{-1}\hat{C}}{Z-1}v\nabla \mathrm{f},
\quad
Z\approx
\begin{cases}
\alpha/2 & \omega \ll \Gamma_{0} \\
1/2 & \omega \gg \Gamma_{0}
\end{cases}
\end{equation*}
Substituting $\mathbf{K}$ into Eq.~\eqref{eq:currentdef}, we derive the conductivity 
\begin{equation}
\sigma = \frac{e^{2}}{\pi h}\int \frac{1}{\alpha} \frac{Z}{1-Z} (-\partial_{\omega}\mathrm{f}_{F}(\omega-\mu))d\omega.
\end{equation}
Let us take the integral for two distinct cases: In the highly doped case, when the system behaves as a Fermi liquid (FL), and, in the charge neutral case, with minimal conductivity (min) when $\mu,T\ll \Gamma_{0}$, the conductivity values are
\begin{equation}
\sigma_\text{FL} = \frac{e^{2}}{\pi h\alpha}
\quad\text{and}\quad
\sigma_\text{min} = \frac{e^{2}}{\pi h}
\end{equation}
correspondingly.
As one can see, the result is consistent with the value predicted before for the order of limits, where the frequency is taken to zero before the scattering rate~\cite{Ludwig1994,Ziegler2007}.
This consistency is intuitively clear when we recall that transport in our case is caused by a static chemical potential gradient.


\section{Conclusions}

The conventional kinetic equation approach for charge neutral Dirac systems generally fails because it requires large Fermi momentum.
We have shown that it is possible to derive an equivalent set of the equations, which operate with the generalized distribution function.
This distribution function has two components: the scalar one, which, in equilibrium, is equivalent to the Fermi distribution, and the vector one, which describes the spin polarizations in Dirac media.
Contrary to conventional distribution functions, the generalized one, similarly to the Green functions, contains also the information about virtual processes encoded in its frequency dependence.
However, unlike in the Green function, its frequency dependence is smooth and without resonances at the mass surface.
This simplifies expansions and calculations.
These novel generalized quantum kinetic equations explicitly demonstrate the influence of spin-momentum locking on transport properties and the presence of off-plane (in 2D) polarization in current.
Solving them, we obtain the correct values for the universal conductivity in 2D Dirac systems, depending on the order of the limits.

\begin{acknowledgements}
B.T.\ acknowledges financial support from the DFG (SPP1666 and SFB1170 "ToCoTronics").
\end{acknowledgements}


\appendix

%

\section{Derivation of the quantum kinetic equation}
\label{apx:qkederiv}

Six Dyson equations (left/right and R/A/K) connect bare Green function $G_{0}$, self-energy $\varSigma$, and full Green function $G$:
\begin{align}
(G_{0}^{-1} - \Sigma^{R/A}) \circ G^{R/A} &=1, 
\label{eq:GRA1} \\
G^{R/A} \circ ( G_{0}^{-1} - \Sigma^{R/A} ) &=1,
\label{eq:GRA2}
\\
(G_{0}^{-1} - \Sigma^{R})\circ G^{K} - \Sigma^{K}\circ G^{A} &=0, 
\label{eq:GK1} \\
G^{K} \circ ( G_{0}^{-1} - \Sigma^{A})- G^{R} \circ \Sigma^{K} &=0. 
\label{eq:GK2}
\end{align}
Here, $(G_{0}^{-1})_{11'} = (i\partial_{t_{1}}-H_{1})\delta_{11'}$.
Equations~\eqref{eq:GRA1}--\eqref{eq:GRA2} are complex conjugate to each other and describe the spectral properties only.
The Keldysh function can be written in the form of an ansatz, which separates the spectral information in $G^{R/A}$ from the occupation numbers~\cite{RammerSmith1986}, namely
\begin{equation}
G^{K} = G^{R} \circ h - h \circ G^{A}.
\end{equation}
Sustituting the ansatz in both left and right equations for the Keldysh component~\eqref{eq:GK1}--\eqref{eq:GK2}, we get
\begin{align}
0 &= h - (G_{0}^{-1} - \Sigma^{R}) \circ h \circ G^{A} - \Sigma^{K} \circ G^{A},
\\
0 &= -h + G^{R}\circ h\circ (G_{0}^{-1} - \Sigma^{A})  - G^{R} \circ \Sigma^{K}.
\end{align}
Multiplying (with convolution) the first equation on the right hand side by $G_{0}^{-1} - \Sigma^{A} =  (G^{A})^{-1}$ and the second  on the left hand side by $(G^{R})^{-1}$, we obtain in both cases the very same equation for the parameter $h$:
\begin{equation}
[G_{0}^{-1}\mathrel{\circ\kern-.91ex\raise-0.3ex\hbox{,}} h]_{-} +  \Sigma^{K} - \Sigma^{R} \circ h + h\circ \Sigma^{A} =0.
\end{equation}
Since the ``lesser'' Green function  $G^{<}_{11'} = -i\langle\psi_{1'}^{+}\psi_{1}\rangle$ can be expressed as $G^{<} = (G^{K}-G^{R}+G^{A})/2$, defining $h = 1-2f$ one gets the expected kinetic equation
\begin{equation}
[G_{0}^{-1}\mathrel{\circ\kern-.91ex\raise-0.3ex\hbox{,}} f]_{-} +  \Sigma^{<} - \Sigma^{R} \circ f + f\circ \Sigma^{A} =0.
\end{equation}
All Green functions, self-energies, and the distribution function $f$ are used in the mixed representation, i.e.\ after partial Fourier transformation
\begin{multline}
G(t,\mathbf{r};\omega,\mathbf{p}) 
= \int d\tau d^{d}\boldsymbol{\rho}\,e^{i\omega \tau - i\mathbf{p}\boldsymbol{\rho}} \times\\
G(\mathbf{r}+\boldsymbol{\rho}/2,\mathbf{r}-\boldsymbol{\rho}/2,t+\tau/2,t-\tau/2).
\end{multline}
In this representation, the convolution transforms into
\begin{equation*}
A\circ B = e^{\frac i2 (
\partial_{t}^{(A)}\partial_{\omega}^{(B)} - \partial_{t}^{(B)}\partial_{\omega}^{(A)}
-
\partial_\mathbf{r}^{(A)}\partial_\mathbf{p}^{(B)} + \partial_\mathbf{r}^{(B)}\partial_\mathbf{p}^{(A)}
)} A B.
\end{equation*}
Since the retarded/advanced Green function is analytical in the upper/lower half-place of the complex variable $\omega$, they both can be expressed through the spectral function $A = i(G^{R} - G^{A})$:
\begin{equation}
G^{R/A}=  \frac i 2 \int \frac{d\omega'}{2\pi} \frac{A}{\omega-\omega'\pm i0}.
\end{equation}
Thus, the correlator can be written in the form
\begin{equation}
\langle \psi_{1'}^{+}\psi_{1}\rangle = i (G^{R} \circ f - f \circ G^{A})= \{A \ccirc f\},
\end{equation}
where the brackets are defined in Eq.~\eqref{eq:Aintdef}.

%
%
%

\section{Calculation of integrals}
\label{apx:sezcalc}


The equation for the relaxation rate $\Gamma$ is obtained within the self-consistent Born approximation.
In the expression $1=\langle 1 \rangle$, the angular brackets are the integral defined in Eq.~\eqref{eq:bracketsdef}.
After the integration over all directions $\mathbf{n}_{\lambda}$, the integral can be reduced to
\begin{equation}
1 = \frac{\alpha}{2}\int_{-E}^{E} \frac{|\xi|d\xi}{(\omega-\xi)^{2}+(\Gamma/2)^{2}}.
\end{equation}
Using dimensionless units $(\Lambda,x,y) = 2(E,\xi,\omega)/\Gamma$, the integral can be calculated as 
\begin{multline}
2\!\int_{-\Lambda}^{\Lambda}\! \frac{|x|dx}{(y-x)^{2}\!+\!1} \!=\!
\log \!\left[\!\frac{(\Lambda\!-\!y)^2\!+\!1}{y^2+1}\frac{(\Lambda\!+\!y)^2\!+\!1}{y^2+1}\right]
\\
+\! 2 y \Bigl[\arctan(\Lambda\!-\!y)\!-\!\arctan(\Lambda\!+\!y) \!+\! 2\arctan(y)\Bigr].
\end{multline}
For $\Lambda\gg z$ and $\Lambda\gg 1$, the expression can be simplified to  
\begin{equation}
\log \frac{\Lambda}{\sqrt{y^2+1}} + y \arctan(y) = \frac1\alpha.
\end{equation}
Considering the limiting cases $y\ll1$ and $y\gg1$, we get
\begin{equation}
\Gamma(\omega) = 
\begin{cases}
\Gamma_{0} & \omega \ll \Gamma_{0}, \\
\pi\omega/\log(2\omega/\Gamma_{0}) & \omega \gg \Gamma_{0},
\end{cases}
\end{equation}
where $\Gamma_0 = 2E e^{-1/\alpha}$. 
Note that, for $\omega \gg \Gamma_{0}$, we obtain
\begin{equation}
\Gamma = \frac{\pi\omega}{\frac1\alpha - \log\frac{E}{\omega}} \approx \pi\alpha\omega,
\end{equation}
a known result for highly doped graphene.

%
%

To calculate the parameter $Z$ defined in Eq.~\eqref{eq:KcalcZdef}, we need to evaluate the integral
\begin{equation}
Z = \frac{\alpha}{2}\int_{-E}^{E}\frac{2\omega \xi +\Gamma^{2}}{4\xi^{2}+\Gamma^{2}} \frac{|\xi|d\xi}{(\omega-\xi)^{2}+(\Gamma/2)^{2}}.
\end{equation}
It is given by the following expression
\begin{equation}
\int_{-\infty}^{\infty}\frac{y x/2 +1}{x^{2}+1} \frac{|x|dx}{(y-x)^{2}+1} = \frac{y^2+1}{y}\arctan y.
\end{equation}
Thus, the factor $Z$ and its values in the limiting cases are equal to
\begin{equation}
Z = \alpha\frac{4\omega^2+\Gamma^{2}}{4\Gamma\omega}\arctan \frac{2\omega}{\Gamma}
\approx
\begin{cases}
\alpha/2 & \omega \ll \Gamma_{0}. \\
1/2 & \omega \gg \Gamma_{0}.
\end{cases}
\end{equation}

\bibliographystyle{apsrev4-2}
\bibliography{diracquantke}

\end{document}